# Statistics of Marginal Wavefunctions as a Real-Space Diagnostic of Quantum Entanglement

Ivan P. Christov*

Physics Department, Sofia University, 1164 Sofia, Bulgaria

email: ivan.christov@phys.uni-sofia.bg

**Abstract**

We present a statistical framework for extracting spatially resolved entanglement directly from an ensemble of marginal (one-body) wavefunctions in Time-Dependent Quantum Monte Carlo (TDQMC). Treating the guide waves as a statistical mixture in Hilbert space, we show that the Gram matrix acts as a covariance operator whose spectrum coincides with the Schmidt spectrum. The associated functional standard deviation closely tracks the von Neumann entanglement entropy both globally and locally via walker partitioning, providing a physically transparent real-space diagnostic of quantum correlations without requiring construction of the full many-body wavefunction. Applications to one-dimensional two-electron bosonic and fermionic systems (helium atom and hydrogen-like molecule) demonstrate excellent agreement with strict conditional-wave results for opposite-spin electrons. For same-spin fermions, TDQMC's statistical treatment of exchange symmetry avoids the unphysical negative local entropies that arise from naive ln(2) subtraction, yielding spatial profiles that remain positive throughout. The method establishes a direct bridge between classical ensemble statistics and quantum entanglement measures, offering a computationally efficient real-space diagnostic tool for mapping the spatial distribution of correlations in many-body systems.

Keywords: spatial entanglement; Gram matrix; quantum Monte Carlo

## 1. Introduction

Statistical descriptions of many-body systems traditionally rely on ensembles of particle configurations, providing access to densities, fluctuations, and correlation functions of observables. In both classical and quantum statistical mechanics, such particle-based statistics form the foundation for understanding equilibrium and nonequilibrium phenomena across physics and chemistry. However, in quantum systems, physical behavior is not determined solely by configuration probabilities but also by phase coherence, interference, and entanglement, which are encoded in the wavefunction itself rather than in particle distributions alone. The distinction between statistics of particles and statistics of waves becomes particularly important in strongly correlated, driven, or nonequilibrium regimes, where quantum coherence plays a central role. Particle ensembles, even when sampled exactly, are intrinsically limited to diagonal observables and classical-like correlations. In contrast, ensembles of wavefunctions defined over physical space retain access to off-diagonal correlations, relative phase information, and mode structure.

In quantum mechanics, the density matrix provides the natural statistical object associated with ensembles of wavefunctions. While often introduced as a bookkeeping device for mixed states or environmental averaging, it can be interpreted more generally as a second-order statistical descriptor of a wave ensemble. Its diagonal elements recover particle densities, whereas its off-diagonal elements encode coherence and spatial correlations. Reduced density matrices, obtained by partial tracing, play a role analogous to marginal covariance matrices in classical statistics, generating all local and subsystem observables. Unlike classical covariance matrices, however, quantum density matrices retain phase-sensitive information and encode entanglement.

In this work, we develop a statistical framework for the ensembles of marginal wavefunctions generated by Time-Dependent Quantum Monte Carlo (TDQMC). We show that the Gram matrix of wavefunction overlaps acts as a covariance operator in Hilbert space whose spectrum fully determines the entanglement structure. We further introduce a spatially resolved analysis by conditioning the ensemble on walker configurations within selected spatial domains. A central result is that the functional standard deviation of the marginal-wave ensemble closely tracks the spatial profile of the von Neumann entanglement entropy, providing a computationally efficient and physically transparent real-space diagnostic of quantum correlations.

## 2. Methods

### A. Classical ensemble statistics

We begin by recalling the standard statistical description of a global ensemble of classical variables (see, e.g., in [1]), which serves as the reference structure for the quantum construction. Consider a set of $M$ realizations of a real-valued random variable (e.g. particle coordinate) $\mathbf{r}^{(k)}$, $k = 1,\ldots,M$ . The sample mean is defined as

$$\overline{\mathbf{r}} = \frac{1}{M}\sum_{k=1}^{M}\mathbf{r}^{(k)} \tag{1}$$

Correlations between components are captured by the covariance matrix

$$\boldsymbol{\Sigma} = \frac{1}{M}\sum_{k=1}^{M}\left(\mathbf{r}^{(k)} - \overline{\mathbf{r}}\right)(\mathbf{r}^{(k)} - \overline{\mathbf{r}})^{T} \tag{2}$$

whose diagonal elements represent variances and off-diagonal elements encode correlations, providing a complete second-order statistical characterization. The total variance measures the spread of the ensemble around the mean

$$\mathrm{Var} = \frac{1}{M}\sum_{k=1}^{M} |\mathbf{r}^{(k)} - \overline{\mathbf{r}}\ |^{2} \tag{3}$$

and can be expressed as the trace of the covariance matrix, $\mathrm{Var} = \mathrm{Tr}\left(\boldsymbol{\Sigma}\right)$. The standard deviation is the square root of the variance $STD = \sqrt{\mathrm{Var}}$ . Thus, diagonalization of $\Sigma$ yields the principal variances $\lambda_k$ , whose sum recovers the total variance. These concepts, mean, variance, covariance matrix, and principal components, form the basis for quantifying statistical spread and effective dimensionality. Below, we show how analogous structures arise naturally for ensembles of marginal wavefunctions, with covariance matrices replaced by Gram matrices (GM) or reduced density matrices (RDM), and principal components replaced by Schmidt decompositions.

### B. Ensembles of functions and functional variance.

We generalize the classical construction to ensembles of functions in Hilbert space. Let $\left\{f^{k}\left(\mathbf{r}\right)\right\}_{k=1}^{M}$ be an ensemble of functions. Statistical properties of such ensembles generalize classical measures for random vectors. First-order statistics describe the mean structure, while second-order statistics describe fluctuations and correlations (see, e.g., [2,3]). The ensemble-averaged function is:

$$\overline{f}\left(\mathbf{r}\right) = \frac{1}{M}\sum_{k=1}^{M}f^{(k)}\left(\mathbf{r}\right) \tag{4}$$

The deviations from the mean are

$$\delta f^{(k)}(\mathbf{r}) = f^{(k)}(\mathbf{r}) - \overline{f}(\mathbf{r}) \tag{5}$$

The functional covariance kernel can be defined as

$$C_{kl}(\mathbf{r}) = \frac{1}{M}\sum_{k=1}^{M}\delta f^{(k)}(\mathbf{r})\delta f^{\dagger(l)}(\mathbf{r}) \tag{6}$$

which is Hermitian and positive semidefinite. The global Hilbert-space variance is

$$\mathrm{Var}[f] = \mathrm{Tr}(C) \tag{7}$$

quantifying the total spread of the ensemble in function space, analogous to the global variance in statistics (Eq.3). The second-order statistical object for the ensemble of functions is the Gram matrix, defined through overlaps between ensemble members

$$G_{kl} = \frac{1}{M}\int f^{(k)}(\mathbf{r}) f^{(l)*}(\mathbf{r})\, d\mathbf{r} \tag{8}$$

which encodes correlations between ensemble members in Hilbert space. Subtracting the squared norm of the mean wavefunction yields the Hilbert-space variance expressed through Gram matrix,

$$\mathrm{Var}[f] = \sum_{k} G_{kk} - \frac{1}{M}\sum_{k,l} G_{kl} \tag{9}$$

which vanishes when all ensemble members coincide and increases with their functional diversity. The corresponding functional standard deviation can be defined as

$$STD_f = \sqrt{\mathrm{Var}[f]} \tag{10}$$

This quantity also vanishes when all functions in the ensemble are identical.

### C. Spatially resolved statistical perspective

The term "spatially resolved statistics" refers to statistical descriptors that depend explicitly on spatial position, obtained by restricting averages or correlations to local neighborhoods or subregions. Such approaches are well established across spatial statistics and image analysis, where local means, variances, and higher moments are computed over spatial domains to characterize heterogeneity and correlations [4–9]. Within the particle context, one may define the local sample mean and local variance of the particles in the spatial domain $\Omega_\alpha$ as $\overline{\mathbf{r}}_\alpha = \sum_{k\in\Omega_\alpha} \mathbf{r}_k / M_\alpha$ where $M_\alpha$ is the number of particles in the domain. The local standard deviation is $STD_\alpha = \sqrt{\mathrm{Var}_\alpha}$ where $\mathrm{Var}_\alpha = \mathrm{Tr}(\boldsymbol{\Sigma}_\alpha)$, $\boldsymbol{\Sigma}_\alpha$ being the local covariance matrix

$$\Sigma_\alpha = \frac{1}{M_\alpha} \sum_{k \in \Omega_\alpha} (\mathbf{r}_k - \overline{\mathbf{r}}_\alpha)(\mathbf{r}_k - \overline{\mathbf{r}}_\alpha)^T \tag{11}$$

Thus, the trace of the local covariance matrix recovers the scalar variance, while its eigenvalues and eigenvectors encode the shape and anisotropy of the local particle cloud. Notice that there is a general theorem in statistics stating that the global variance of a dataset can be expressed as the sum of the mean of local variances and the variance of local means (variance decomposition formula [1]).

However, the direct transfer of functional statistics to restricted spatial domains is not straightforward, since it depends on the concrete model under consideration. Such a transfer has been attempted in the past in the field of quantum entanglement. In other work, spatially resolved perspectives on entanglement have been explored through constructions such as “entanglement contours”, which aim to decompose the total entanglement entropy of a subsystem into position-dependent contributions [10,11] or using localized orbitals [12]. These approaches provide physically motivated ways to associate entanglement with real-space degrees of freedom where the lack of unique local definitions is well recognized and reflects the fundamentally nonlocal nature of quantum entanglement. Other research investigates the localization of the entanglement of two interacting distinguishable particles in the configuration space by sampling a preliminary calculated reduced density matrix for known wavefunctions over a uniform grid and next calculating the von Neumann entropy [13].

Here we show another approach to functional statistics in the case of an ensemble of wave functions generated by the TDQMC method [14-16], which offers a particularly transparent framework in which both particle and wave statistics coexist. In TDQMC, ensembles of stochastic trajectories are coupled to a set of wavefunctions that evolve concurrently in physical space. This dual representation allows oneto extract conventional particle-based observables alongside reduced density matrices and wave-ensemble correlation measures within the same calculation. As a result, TDQMC provides a natural platform for comparing particle and wave statistical descriptions on equal footing and for identifying which physical information is uniquely accessible through wave-based statistics.

### D. TDQMC as a statistical framework for wave ensembles

We consider a many-body quantum system described through a particle–wave dual representation, in which an ensemble of particle trajectories is coupled to a corresponding ensemble of wavefunctions. Each ensemble exhibits its own statistical properties, which may encode correlation features of the underlying quantum system. Such a structure arises naturally in quantum simulations where the many-body wavefunction $\Psi(\mathbf{r}_1, \mathbf{r}_2, ..., \mathbf{r}_N)$ is conditioned on particle trajectories $\mathbf{r}_i^k(t)$, as in conventional quantum Monte Carlo methods using analytical

Slater–Jastrow forms (conditional wavefunctions) [17], or as in the TDQMC framework, where a set of one-body wavefunctions $\left\{\varphi^k\left(\mathbf{r},t\right)\right\}_{k=1}^{M}$ and particle trajectories ("walkers") $\left\{\mathbf{r}^k\left(t\right)\right\}_{k=1}^{M}$ evolve concurrently in physical space-time where each walker samples the distribution given by its guide wave [14–16]. This dual representation provides access to both particle-based statistics, through the distribution of walker positions, and wave-based statistics, through ensembles of wavefunctions. Importantly, these two ensembles are not independent: they are coupled through effective interaction potentials constructed via Monte Carlo convolution over the walker distribution [15]. In this work, we refer to the correlated one-body wavefunctions, defined either as strict conditional wavefunctions $\Psi\left(\mathbf{r}_1\left(t\right),...,\mathbf{r},...,\mathbf{r}_N\left(t\right)\right)$ or as TDQMC guide waves $\varphi_i^k\left(\mathbf{r},t\right)$, as *marginal waves*. Notice that the individual marginal waves do not carry independent ontological meaning; rather, their statistical properties as an ensemble encode physically relevant information. For ground-state calculations, these are often real functions.

For a system of $N$ electrons, the TDQMC guide wavefunctions for the *i*-th electron $\varphi_i^k\left(\mathbf{r}_i,t\right)$ *i=1,…,N,* evolve according to single-particle Schrödinger-like equations [18]

$$i\hbar\frac{\partial}{\partial t}\varphi_i^k(\mathbf{r}_i,t)=\left[-\frac{\hbar^2}{2m}\nabla_i^2+V_{en}(\mathbf{r}_i)+V_{eff}^k\left(\mathbf{r}_i,t\right)\right.$$
$$\left.-\sum_{j\neq i}^{N}\delta_{s_i,s_j}\int d\mathbf{r}_j V_{ee}(\mathbf{r}_i,\mathbf{r}_j)\varphi_i^k(\mathbf{r}_j,t)\varphi_j^{k*}(\mathbf{r}_j,t)\varphi_j^k(\mathbf{r}_i,t)/\varphi_i^k(\mathbf{r}_i,t)\right]\varphi_i^k(\mathbf{r}_i,t) \tag{12}$$

where $V_{en}\left(\mathbf{r}\right)$ is the nuclear potential and the indices $s_i,s_j$ denote the spins of the *i*-th and the *j*-th electrons. The effective electron-electron interaction includes a normalized Monte-Carlo convolution of the interparticle potential $V_{ee}\left(\mathbf{r}_i,\mathbf{r}_j\right)$

$$V_{eff}^k\left(\mathbf{r}_i,t\right)=\sum_{j\neq i}^{N}\frac{\sum_l^M V_{ee}\left[\mathbf{r}_i,\mathbf{r}_j^l(t)\right]K\left[\mathbf{r}_j^l(t)-\mathbf{r}_j^k(t),\sigma_j\right]}{\sum_l^M K\left[\mathbf{r}_j^l(t)-\mathbf{r}_j^k(t),\sigma_j\right]} \tag{13}$$

and for parallel-spin electrons the guide waves satisfy orthonormality $\int\varphi_i^k(\mathbf{r},t)\varphi_j^{k*}(\mathbf{r},t)d\mathbf{r}=\delta_{i,j}$; *i=1,…,N, k=1,…,M.*

The (Gaussian) convolution kernel $K\left[\mathbf{r}_j,\mathbf{r}_j^k(t),\sigma_j\right]$ in Eq.13 and its width $\sigma_j$ introduce a variationally controlled nonlocality, allowing each guiding wavefunction to incorporate information about the ensemble distribution of the other particles. This framework interpolates between pairwise interactions among walkers ($\sigma_j\to 0$) [14] and the mean-field (Hartree–Fock) approximation ($\sigma_j \to \infty$). Crucially, this "communication" occurs through the wavefunctions rather than through direct particle–particle forces, a feature that has no classical analog. Exchange symmetry is treated differently for bosons and fermions. For bosons, TDQMC does not explicitly enforce symmetry at the level of a many-body wavefunction; instead, symmetry is preserved statistically by sampling permutation-invariant configurations. As a result, all permutation-invariant observables constructed from TDQMC are exactly symmetric under particle exchange. For fermions, exchange symmetry is incorporated through anti-symmetrized constructions such as Slater determinants at the level of the guide waves. The TDQMC equations can be generalized to include spin by introducing spinor wavefunctions, leading to coupled equations for spin components, with or without external fields.

### E. Marginal-wave ensembles and Hilbert-space statistics

The ensemble of marginal waves provides a representation of the many-body state in Hilbert space, where objects such as the Gram matrix and the reduced density matrix reveal the structure of correlations and entanglement through their spectra. In this sense, the marginal-wave ensemble forms a bridge between the Hilbert-space description of entanglement and the real-space structure of electronic correlations and is related to the geometry of the state space [19]. We define the *global* Gram matrix for a given electron as (see Eq.8)

$$G_{kl}(t) = \frac{1}{M}\int \varphi^k(\mathbf{r},t)\varphi^{l*}(\mathbf{r},t)\mathrm{d}\mathbf{r} \tag{14}$$

This matrix captures the overlap geometry of the marginal-wave ensemble and acts as a covariance operator in Hilbert space. It is Hermitian and positive semidefinite. Its rank, determined by its eigenvalue spectrum, reflects the effective dimensionality of the ensemble: a rank of one corresponds to identical marginal waves (uncorrelated case), while higher rank indicates increasing diversity and correlation. The spread of the ensemble is quantified by the global Hilbert-space variance (see Eq.9)

$$\mathrm{Var}_{\mathcal{H}}(t) = 1 - \frac{1}{M}\sum_{k,l} G_{kl}(t) = 1 - \int \left|\overline{\varphi}(\mathbf{r},t)\right|^2 d\mathbf{r} \tag{15}$$

and the corresponding global functional standard deviation

$$STD_{\mathcal{H}}(t) = \sqrt{\mathrm{Var}_{\mathcal{H}}(t)} \tag{16}$$

This quantity generalizes classical variance to ensembles of wavefunctions, where the Gram matrix encodes the full Hilbert-space geometry. In the Hartree–Fock limit, all marginal waves coincide and the standard deviation vanishes, despite maximal spatial overlap. Nonzero values indicate correlation-induced diversity. The same ensemble defines a global reduced density matrix

$$\rho(\mathbf{r},\mathbf{r}',t) = \frac{1}{M}\sum_{k=1}^{M} \varphi^{k*}(\mathbf{r},t)\varphi^{k}(\mathbf{r}',t) \qquad (17)$$

which is Hermitian and positive semidefinite, with unit trace. In this way, reduced density matrices and entanglement measures arise naturally as second-order statistical descriptors of the TDQMC ensemble, without explicit construction of the full many-body wavefunction. Therefore, the density matrix (Eq.17) may be used to measure quantum correlations where it is regarded as the density operator of the statistical ensemble of marginal waves, rather than as the exact reduced density matrix obtained by tracing the many-body wavefunction. By introducing a linear operator $\hat{A}$ such that its action on a general vector $\mathbf{c} = (c_1,\ldots,c_M)$ is given by $\hat{A}\mathbf{c} = 1/\sqrt{M}\sum_{k=1}^{M} c_k \varphi^k$, it is seen that the Gram matrix and the reduced density matrix can be written as $G = \hat{A}^{\dagger}\hat{A}$, $\hat{\rho} = \hat{A}\hat{A}^{\dagger}$. Then, it follows directly from the singular value decomposition of $\hat{A}$ that the GM and RDM share the same nonzero eigenvalues (spectrum). In other words, the entanglement spectrum of the ensemble state depends on its statistical measures determined by the functional standard deviation and the overlap geometry. In a broader sense, the Gram matrix constructed from the TDQMC marginal-wave ensemble is mathematically equivalent to the kernel matrix used in kernel principal component analysis [20]. Its eigenvalues therefore quantify the effective dimensionality of the ensemble in Hilbert space. In this sense, the Schmidt spectrum can be interpreted as the principal-component spectrum of the marginal-wave ensemble.

One may also construct cross-electron Gram matrices $G_{kl}^{cross}(t) = \int \varphi_i^k(\mathbf{r},t)\varphi_j^{l*}(\mathbf{r},t)d\mathbf{r} \big/ M$ between different electrons $i$ and $j$, providing information about inter-electron alignment and structural correlations. In general, the Gram matrix has found numerous applications in quantum physics including quantum dynamics [21], quantum coherence [22], and entanglement [23].

### F. Local Gram matrix and local reduced density matrix

We exploit the Gram matrix constructed from the marginal wavefunctions to quantify their Hilbert-space variance and effective dimensionality. While covariance-matrix participation ratios and effective ranks are well known in linear algebra and statistics [24], their use as a spatially resolved diagnostic of entanglement and coherence in real-space wave ensembles appears to be unexplored. We assume that localization occurs entirely in walker configuration space. Specifically, we introduce spatial partitioning in the form of a set of domains $\Omega_\alpha$, $\alpha = 1,2,\ldots$ which act exclusively on the walker ensemble and do not restrict the support of the associated marginal waves, which remain global functions. Therefore, spatial locality enters solely through the selection of the corresponding marginal waves contributing to the local Gram and reduced density matrices. In a discrete representation, the *local* Gram matrix of the guide-wave ensemble for a given electron is defined as a matrix with elements

$$G_{kl}^{(\alpha)}(t) = \frac{1}{M_\alpha}\int \varphi^k(\mathbf{r},t)\varphi^{l*}(\mathbf{r},t)\,d\mathbf{r}, \qquad k,l \in \Omega_\alpha \tag{18}$$

and is constructed from the marginal wave ensemble associated with the walkers located within the spatial domain $\Omega_\alpha$, while the integrals defining the overlaps extend over the full spatial domain. If the domain size tends to zero, only a single walker remains within the domain. In this limit, the local Gram matrix $G^{(\alpha)} = \left[\langle \psi_k \,|\psi_k\rangle\right]$ reduces to a rank-one matrix, indicating that the Hilbert-space dimensionality collapses. Therefore, a finite domain size is required to reveal the geometry of the ensemble and to meaningfully quantify correlations.

The *local* Hilbert-space variance, defined analogously to Eq. 15, is given by

$$\mathrm{Var}_{\mathcal{H}}^{(\alpha)}(t) = 1 - \frac{1}{M}\sum_{k,l} G_{kl}^{(\alpha)}(t) \tag{19}$$

and provides a sensitive diagnostic of spatially resolved entanglement, reproducing its qualitative spatial structure. We find that the local Hilbert-space standard deviation of marginal wavefunctions closely tracks the spatial profile of the entanglement entropy. This observation indicates that local entanglement enhancement originates from increased diversity of marginal waves rather than from their overlap alone.

### G. Local reduced density matrix

For each domain $\Omega_\alpha$ in configuration space, a reduced density matrix can be constructed by forming an ensemble average over the marginal waves associated with the walkers belonging to that domain. This defines a *local* reduced density matrix representing the quantum state conditioned on the $i$-th electron being located within $\Omega_\alpha$

$$\rho_i^{(\alpha)}(\mathbf{r},\mathbf{r}',t) = \frac{1}{M_\alpha}\sum_{k\in\Omega_\alpha}\varphi_i^{k*}(\mathbf{r},t)\varphi_i^{k}(\mathbf{r}',t) \tag{20}$$

which is normalized to unit trace. Entanglement measures computed from this reduced density matrix therefore quantify the entanglement between the $i$-th electron localized in $\Omega_\alpha$ and the rest of the system. By repeating this construction over all spatial domains, a spatially resolved entanglement profile is obtained. Variations in the local entanglement reflect changes in the diversity and structure of the guide-wave ensemble within each domain, rather than variations in local density alone. It should be emphasized that, although we refer to domains in configuration space, in practical implementations these domains are typically defined along physical spatial axes (see Figure 1), allowing efficient computation of the local density matrix.

The local entanglement is quantified using entropic measures derived from the reduced density matrix, such as the von Neumann entropy (e.g. in [25])

$$S_i^{(\alpha)} = -\mathrm{Tr}\left(\rho_i^{(\alpha)} \ln \rho_i^{(\alpha)}\right) \tag{21}$$

which measures the entanglement between the *i*-th electron localized in $\Omega_\alpha$ and the rest of the system [26,27]. Domains where the guide waves are nearly identical correspond to weak entanglement, while regions where the marginal-wave ensemble spans a larger effective Hilbert space give rise to enhanced local entanglement. In the limit where the domains collectively cover the entire configuration space of the electron, the spatially resolved entanglement profile provides a decomposition of the global entanglement into spatial contributions, offering a real-space picture of how quantum correlations are distributed across the system.

More generally, the Hilbert-space variance of the marginal-wave ensemble is directly related to the deviation of the reduced density matrix from a rank-one projector. To characterize the effective dimensionality of the marginal-wave ensemble within each domain, we define an effective Schmidt number:

$$K_{\mathrm{eff}}^{(\alpha)} = \frac{1}{\mathrm{Tr}\,(G^{(\alpha)})^2} \tag{22}$$

This quantity represents the number of locally significant Schmidt modes and serves as a clear diagnostic of cooperative delocalization. The effective Schmidt number can be estimated directly from the eigenvalues of the local Gram matrix of marginal wavefunctions. An increase in the number of linearly independent marginal waves contributing within a spatial region corresponds to an increase in the local Schmidt rank and underlies the observed enhancement of local entropy.

The above formalism can be straightforwardly generalized to systems with spin by introducing spin-resolved Gram matrices and corresponding reduced density matrices, consistent with the spinor extension of the TDQMC equations.

### H. Spectral representation

The close relationship between the Gram matrix and the reduced density matrix constructed from the same marginal-wave ensemble, namely, that they share the same nonzero eigenvalues, can be directly extended to local quantities, provided proper normalization within each domain is ensured.

Let $\left\{\lambda_m^{(\alpha)}\right\}$ be the eigenvalues of the local Gram matrix (or equivalently the local reduced density matrix). From normalization of the marginal waves, we have

$$\sum_m \lambda_m^{(\alpha)} = 1 \tag{23}$$

and therefore:

$$\sum_{k,l} \left| G_{kl}^{(\alpha)} \right|^2 = \sum_m \lambda_m^{(\alpha)2} = \mathrm{Tr}\left( \rho^{(\alpha)2} \right) \tag{24}$$

Using these eigenvalues, the local functional standard deviation can be expressed as

$$\left( STD_\alpha \right)^2 = 1 - \frac{1}{M} \sum_{k,l} G_{kl}^{(\alpha)} \tag{25}$$

while the local von Neumann entropy is given by

$$S^{(\alpha)} = -\sum_m \lambda_m^{(\alpha)} \ln \lambda_m^{(\alpha)} \tag{26}$$

In addition, the local linear entropy, which provides a second-order measure of statistical mixing associated with the reduced density matrix, is defined as

$$S_L^{(\alpha)} = 1 - \mathrm{Tr}\left( \rho^{(\alpha)2} \right) = 1 - \sum_{k,l} \left| G_{kl}^{(\alpha)} \right|^2 \tag{27}$$

From a physical perspective, if all guide waves within a domain are identical (as in the Hartree–Fock limit), the Gram matrix has rank one, the eigenvalue spectrum collapses to a single nonzero value, and both the entropy and the functional standard deviation vanish. Interaction-driven divergence of the marginal-wave ensemble increases the effective rank of the Gram matrix and produces multiple nonzero local Schmidt coefficients. This broadens the eigenvalue spectrum, and both the entropy and the functional standard deviation increase. Thus, the functional standard deviation quantifies Hilbert-space dispersion in the same way that classical standard deviation quantifies spatial dispersion in a particle ensemble.

## 3. Results and Discussion

### A. Local ensemble statistics under spatial partitioning

To obtain a spatially resolved characterization of entanglement within the TDQMC framework, the walker ensemble corresponding to a selected electron is partitioned into spatial domains $\Omega_\alpha$ along its physical coordinate. As illustrated in Figure 1 for two one-dimensional electrons at a finite separation, the partitioning is performed by dividing the one electron's coordinate into non-overlapping strips of width $\Delta x = x_{right} - x_{left}$, each strip defining a domain $\Omega_\alpha$ that collects all walkers whose instantaneous positions fall within that interval, without imposing restrictions in the remaining directions.

For each domain, ensemble functional statistics are computed using only the marginal waves associated with walkers located within that region. This construction defines a local functional mean and a local functional standard deviation, characterizing the spread of the ensemble conditioned on the electron being located within $\Omega_\alpha$. In this way, the local functional standard deviation measures the internal variability of the wave ensemble within each spatial domain, independently of the total density or the global spread of the walker distribution.

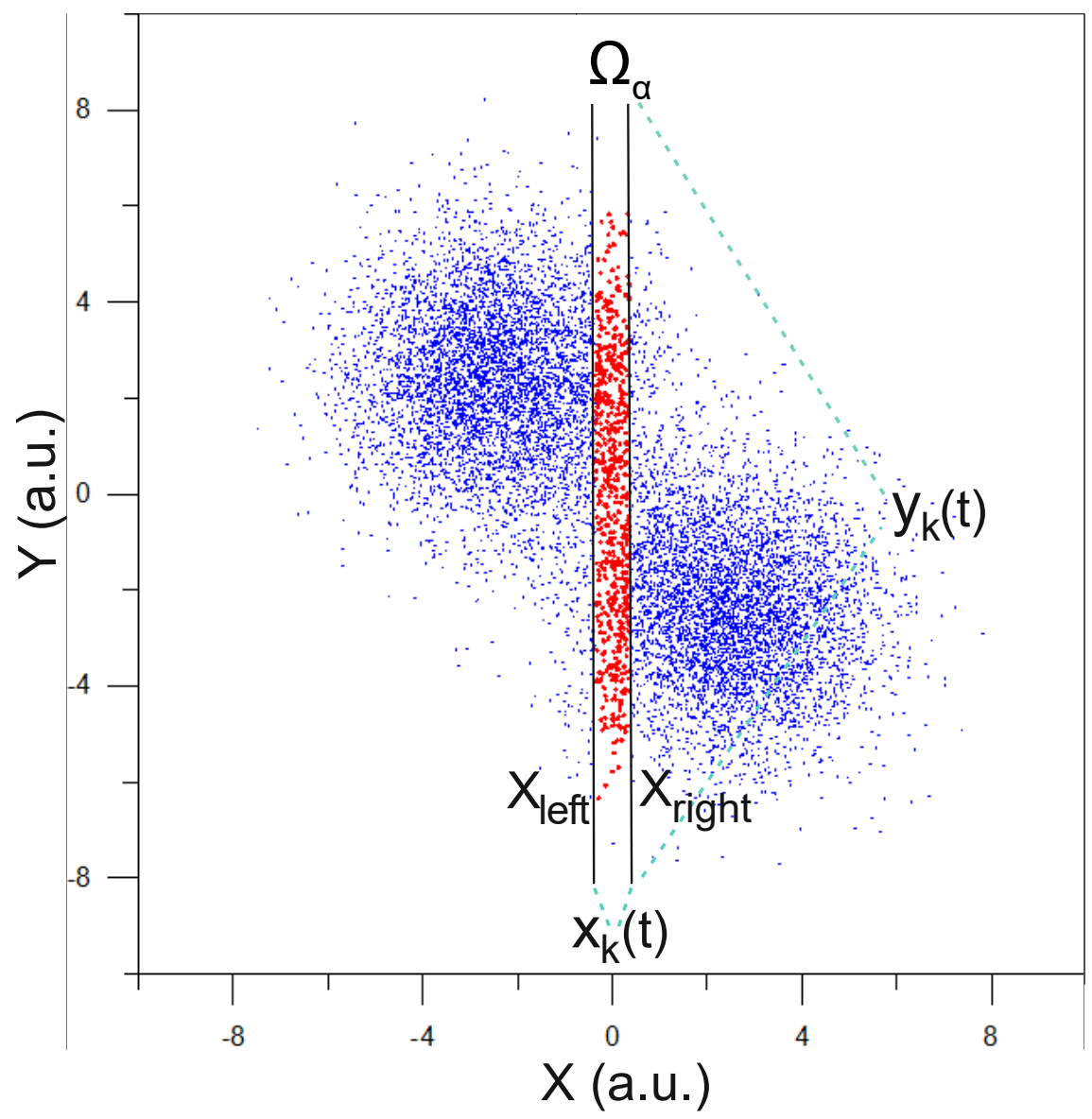

**Figure 1**. Walker distribution in configuration space for a one-dimensional hydrogen-like molecule. The central red strip defines the spatial domain $\Omega_\alpha$ used to compute the local Gram matrix (GM) and reduced density matrix (RDM). This domain constrains only the walkers; the associated guide waves retain unrestricted spatial support over the full domain. Green dashed lines indicate the ranges of walker coordinates $x$ and $y$ within the selected domain.

For the configuration shown in Figure 1, at a given time $t$, the electron cloud is represented by a finite ensemble of walkers $\{\mathbf{r}_k(t)\}_{k=1}^{M}$, $\mathbf{r}_k(t)=\{x_k(t),y_k(t)\}$, and the domain $\Omega_\alpha=\left\{(x_k,y_k):x_k\in\left[X_{left},X_{right}\right]\right\}$, is defined by the subset of walkers indicated by the red points (the vertical strip in Figure 1).

In order to find the system ground state, in all simulations presented below, the total number of walkers and guide waves for a single electron was $M=25600$. Imaginary-time propagation was performed with a time step $\Delta\tau=0.02$ a.u. for 400 steps. The nonlocal interaction kernel $K$ in Eq.13 was taken to be Gaussian, with its width determined by the energy-minimization procedure described in Refs. [16,18]. The resulting optimal widths were $\sigma \approx 1$ a.u. for opposite-spin electrons and $\sigma \approx 1.5$ a.u. for parallel-spin electrons. The spatial domains considered in the local entropy analysis extended over 16 a.u. for the parallel-spin systems and 10 a.u. for the opposite-spin systems. The spatial domain was partitioned into 11 strips, corresponding to strip widths of approximately 1.45 a.u. for the fermionic systems and 0.91 a.u. for the bosonic systems. The number of strips was chosen as a compromise between spatial resolution and statistical stability. Increasing the number of strips reveals finer spatial structure in both the functional standard deviation and local entropy profiles but also amplifies statistical fluctuations. The overall localization patterns and average trends remain qualitatively unchanged. In the fermionic case, the chosen partition additionally resolves the exchange-induced cusp region, facilitating analysis of its influence on the local ln(2)-corrected entropy.

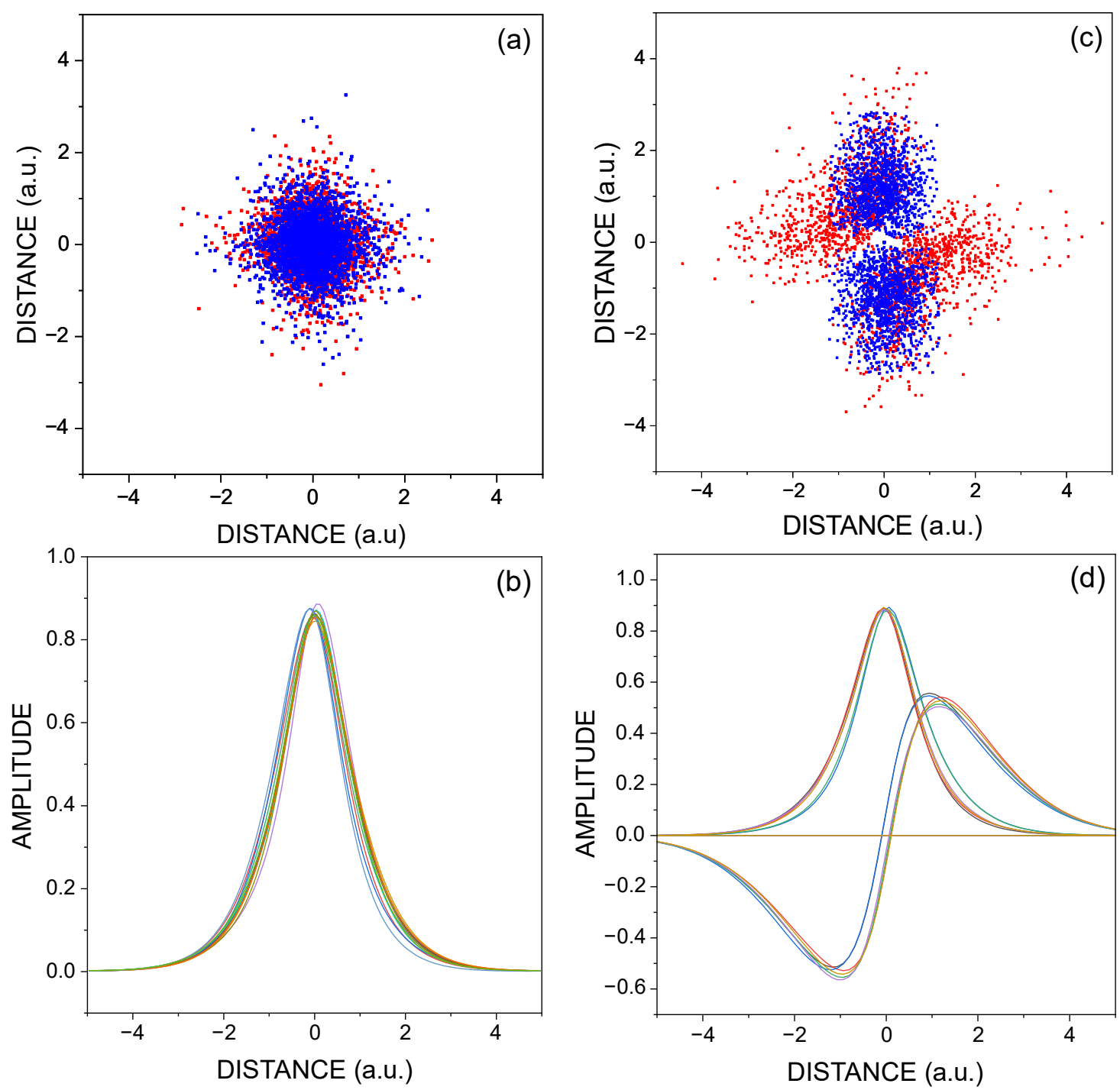

**Figure 2.** Walker distributions in configuration space and corresponding guide waves for the one-dimensional helium atom after imaginary-time propagation to the ground state. Walker distributions for opposite-spin (parahelium) (a) and parallel-spin (orthohelium) (c) electrons. Red dots: exact results from the numerical solution of the two-dimensional Schrödinger equation. Blue dots: TDQMC results. (b,d) Representative TDQMC guide waves.

Although entanglement is fundamentally a nonlocal quantity, the local reduced density matrix $\rho^{(\alpha)}$ constructed using restricted walker domains $\Omega_\alpha$ corresponds to a position-conditioned quantum state. This construction can be interpreted as performing a *weak measurement* of the electron's position followed by post-selection. Therefore, it provides a well-defined and operationally meaningful measure of local entanglement within the TDQMC framework.

### B. One-Dimensional Atom

We shall consider statistical measures for ensembles of guide waves for a 1D helium atom with two opposite-spin electrons (parahelium) and two parallel-spin electrons (orthohelium). Although simplified, this approach allows for a direct comparison between the numerically exact results using 1D strict conditional wavefunctions and the TDQMC ensemble of guide waves. Figure 2(a) shows a typical configuration-space distribution of walkers for parahelium, while Figure 2(b) displays a representative subset of the corresponding TDQMC guide waves. These results are obtained after imaginary-time propagation of both the TDQMC equations and the exact two-dimensional Schrödinger equation until convergence to the stationary state is achieved, with soft electron–nuclear and electron–electron interactions fully included (see Eqs. 12, 13, and [16]).

Figure 2(c) presents the configuration-space distribution for orthohelium as predicted by TDQMC (blue dots), compared with the Monte Carlo sampling of the exact distribution obtained from the two-dimensional Schrödinger equation (red dots). As seen in Figure 2(c), because TDQMC operates in physical space rather than configuration space, the resulting walker distribution is closer to a product-state-like distribution and does not reproduce the pronounced cusp along the diagonal characteristic of the exact distribution. Figure 2(d) shows several samples of TDQMC guide waves corresponding to both ground and excited fermionic states, which remain mutually orthogonal at each instant in time.

It should be noted that TDQMC allows electrons to be treated either as distinguishable or identical particles. At each time step, a local Slater determinant can be constructed numerically from the guide waves associated with walkers within a given domain to enforce the Pauli exclusion principle and to compute entanglement entropy for identical particles. However, such local constructions do not generate the global cusp observed in the exact configuration-space distribution such as that in Figure 2(c), red dots. Whenever the global many-body probability distribution in configuration space is required, for example, for calculating the ground state energy as a function of parameter $\sigma_j$ in Eq.13, a separate auxiliary set of walkers can be introduced to sample an anti-symmetric wavefunction (Slater determinant) built from guide waves [18,28].

Figure 3(a,b) presents spatially resolved second-order statistical measures for parahelium, computed over 11 spatial strips along the coordinate axes of Figure 2(a). The local functional standard deviation (Eq.19 and Eq.16) and the local von Neumann entropy (Eq. 21) obtained from TDQMC guide waves are shown by blue curves, while the corresponding results from strict conditional wavefunctions are shown in red. It is observed that the spatial profiles predicted by TDQMC closely match the exact results within statistical uncertainty. More importantly, the spatial dependence of the local entanglement entropy in Figure 3(b) closely mirrors that of the local

functional standard deviation in Figure 3(a). The peak of the entropy in Figure 3(b), located at the central strip near $x = 0$, approaches the exact global entanglement entropy (indicated by the green marker in Figure 3(b)).

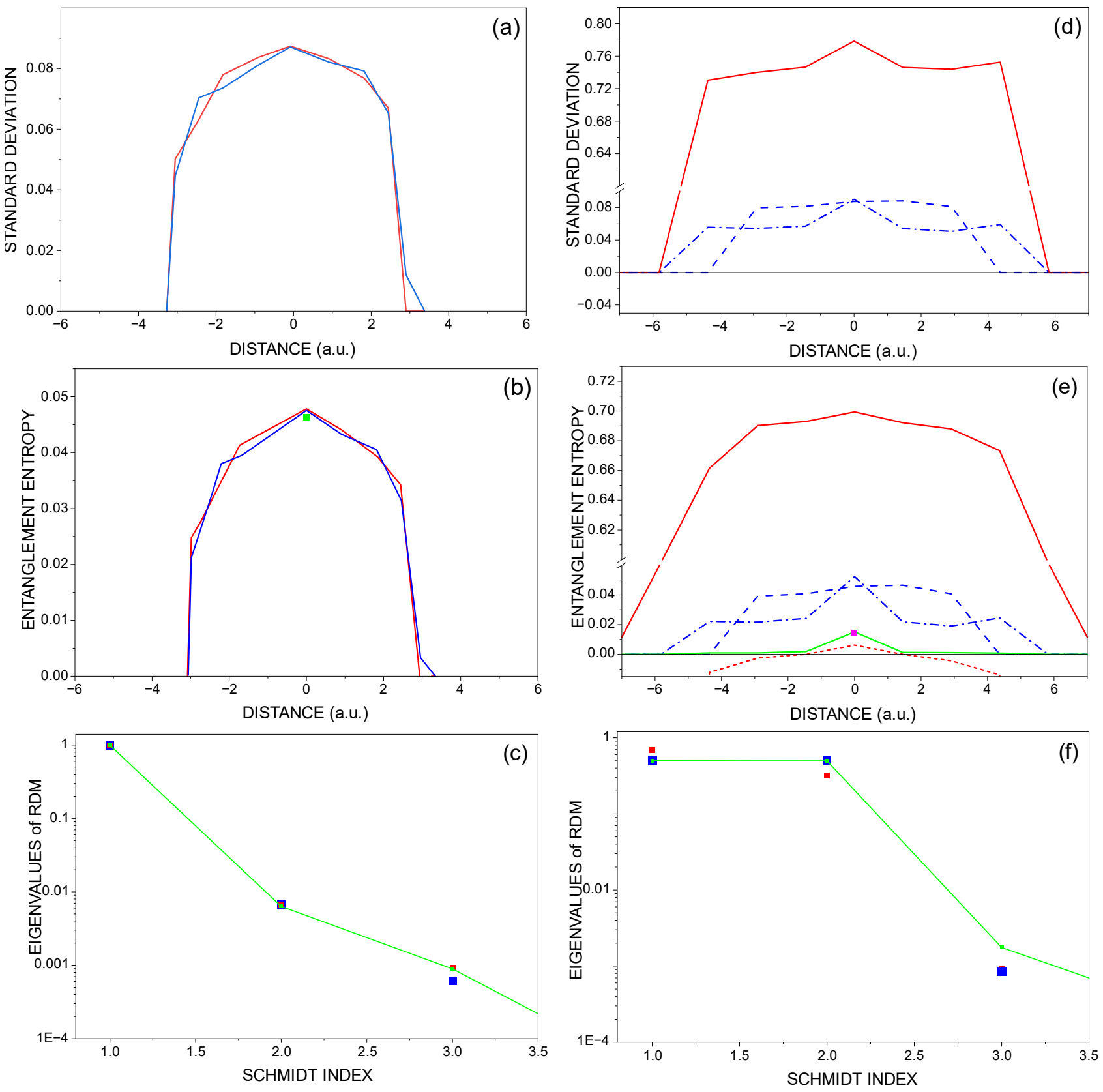

**Figure 3.** Spatially resolved functional standard deviation (a,d), von Neumann entanglement entropy (b,e), and Schmidt spectra (c,f) for the one-dimensional helium atom. Panels (a–c) show opposite-spin (parahelium) electrons and panels (d–f) show parallel-spin (orthohelium) electrons. Red lines/markers: exact conditional-wave results. Blue dashed and dash-dot lines/blue markers: TDQMC guide-wave results for the ground state (dash) and for the fermionic excited state (dash-dot) for distinguishable fermions. Green marker in (b) and magenta marker in (e): exact global entanglement entropy. In panel (e), the dashed red line shows the ln(2)-corrected results for identical fermions from the exact calculation while the green line shows the local entropy obtained from Slater determinants constructed from TDQMC guide waves for identical fermions, after ln(2) correction. Panels (c,f) compare local Schmidt spectra from the central strip of strict conditional waves and TDQMC guide waves calculation (red and blue markers, respectively) with the exact global spectrum (green line with markers).

These observations provide direct evidence that the entanglement entropy can be interpreted as a measure of disorder within the marginal-wave ensemble. While the variance quantifies the spread of marginal wavefunctions around their mean, the von Neumann entropy reflects the effective number of contributing Schmidt components. Consequently, the functional standard deviation may exceed the von Neumann entropy, as observed in Figures 3(a,b), and may decay more slowly in weakly correlated regimes.

The consistency between TDQMC and exact conditional-wave statistics is further confirmed by the Schmidt spectra shown in Figure 3(c). The exact global eigenvalues of the reduced density matrix (green line with markers) are compared with those obtained from the central strip using both strict conditional waves (red markers) and TDQMC guide waves (blue markers). The leading eigenvalues exhibit close agreement, with the remaining discrepancies attributable to statistical noise.

A corresponding analysis for orthohelium is presented in Figures 3(d–f). One of the main challenges in quantifying entanglement for fermions is the contribution associated with the antisymmetric nodal structure (cusp) in configuration space, as evident in Figure 2(c), red dots. The Pauli principle enforces antisymmetry, which for two electrons introduces a global entropy contribution of magnitude ln(2) that is not associated with interaction-induced correlations [29-31]. Although this issue remains under discussion, it is commonly assumed that this geometric contribution should be subtracted to isolate the true correlation entropy. However, as shown by red dashed line in Figure 3(e), a naive subtraction of ln(2) from the exact local entropy leads to unphysical negative values over a large spatial region. The only positive contribution remains in the central strip around x=0, which covers the area where the global cusp for the true 2D walker distribution is located. These results indicate that, although subtraction of the ln(2) contribution is commonly used to isolate interaction-induced correlations, its direct application to spatially resolved entropies may lead to inconsistencies, such as negative local values.

In contrast, in TDQMC, where electrons can be treated as distinguishable or as identical particles, the resulting local entanglement entropy for identical particles remains much lower but positive everywhere after ln(2) correction (green line in Figure 3(e)). The spatial profiles of the entanglement entropy and the functional standard deviation in Figures 3(d), 3(e) are very similar. For identical electrons the local entropy exhibits a central peak close to the exact global value (magenta mark in Figure 3(e)), after ln(2) correction. The Schmidt spectra in Figure 3(f) further confirm that the local reduced density matrix constructed near the cusp in the strict conditional wave solution (red markers) does not reliably reflect the global entanglement, in contrast to the TDQMC guide wave case (blue markers), where no cusp is present. This indicates that the TDQMC statistical treatment of exchange avoids the negative local entropies produced by direct local ln(2) subtraction and yields a positive, well-defined description of spatially resolved fermionic entanglement. It should be noted that for a larger number of contiguous spatial domains, the similarity between the spatial localization patterns of the functional standard deviation and the von Neumann entropy can be quantified using other statistical measures, such as the Pearson correlation coefficient.

### C. One-Dimensional Molecule

To further investigate marginal-wave statistics in extended systems, we consider a model one-dimensional hydrogen-like molecule consisting of two atoms separated by a distance of 3 a.u., for both opposite-spin (parahydrogen) and parallel-spin (orthohydrogen) configurations. Figure 4(a,c) shows that, in both cases, the walker distributions and corresponding guide-wave ensembles split into two sub-ensembles localized near the nuclei. Figure 4(b,d) shows representative samples of the corresponding TDQMC guide waves, including both ground and excited fermionic states for the orthohydrogen molecule.

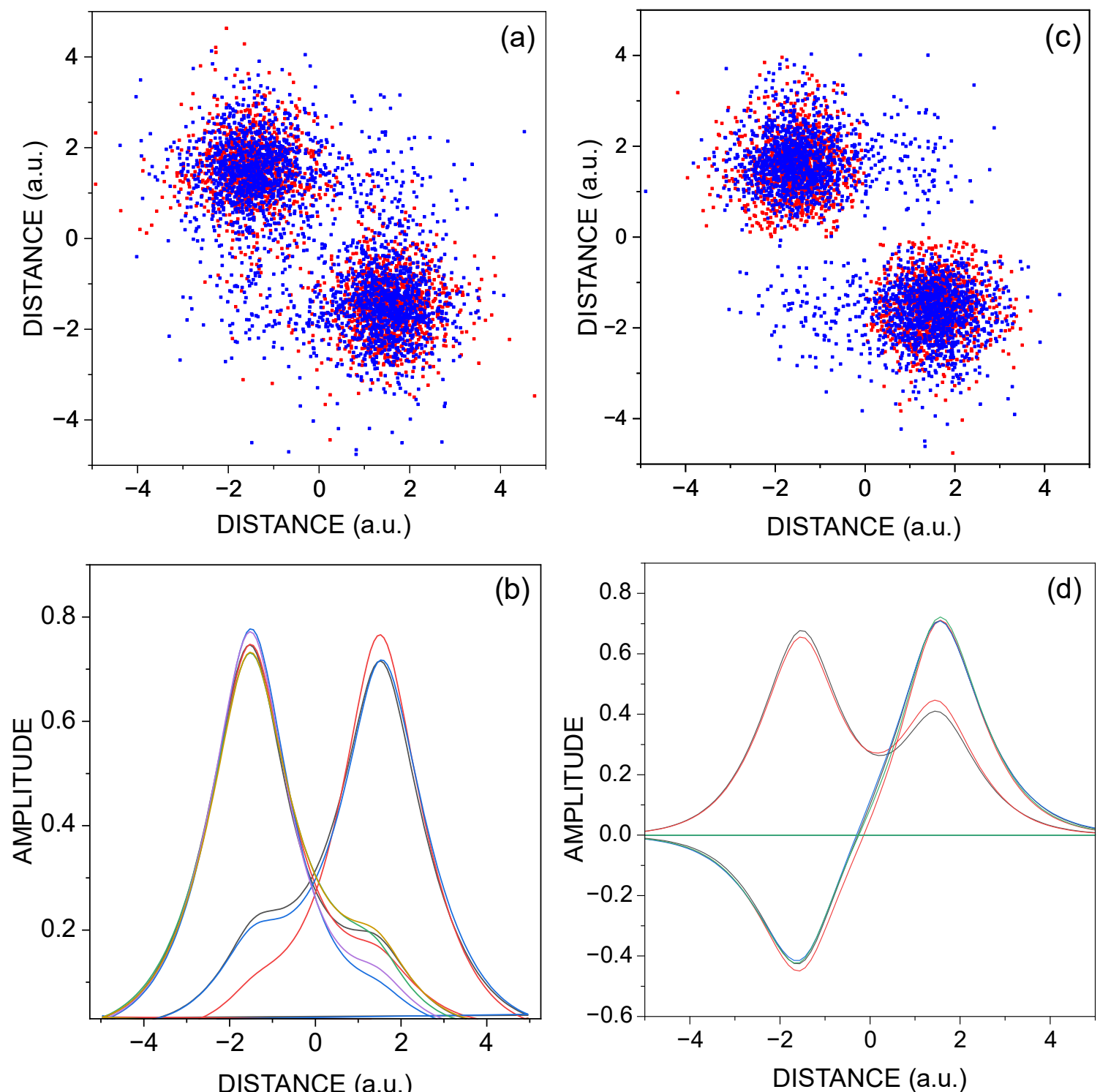

**Figure 4.** Walker distributions in configuration space and corresponding guide waves for the one-dimensional hydrogen-like molecule (internuclear distance 3 a.u.) after imaginary-time propagation to the ground state. (a,c) Walker distributions for opposite-spin (parahydrogen) and parallel-spin (orthohydrogen) electrons. Red dots: exact results. Blue dots: TDQMC results. (b,d) Representative TDQMC guide waves for ground (b) and excited fermionic (d) states.

Figure 5(a,b) shows the spatially resolved functional standard deviation and entanglement entropy for the parahydrogen molecule. Both quantities exhibit a pronounced peak in the central region (around $x = 0$), consistent with previous studies [26,27]. As in the atomic case in Figure 3, the spatial profiles of the functional standard deviation and entropy are remarkably similar, reinforcing the interpretation of entropy as a measure of ensemble diversity. The exact global entropy (green marker in Figure 5(b)) lies close to the peak of the local entropy distribution, indicating once again that the central strip captures the dominant contribution to entanglement (cf. Figure 3(b)). The Schmidt spectrum for parahydrogen (Figure 5(c)) shows good agreement between TDQMC and exact results for the leading eigenvalues.

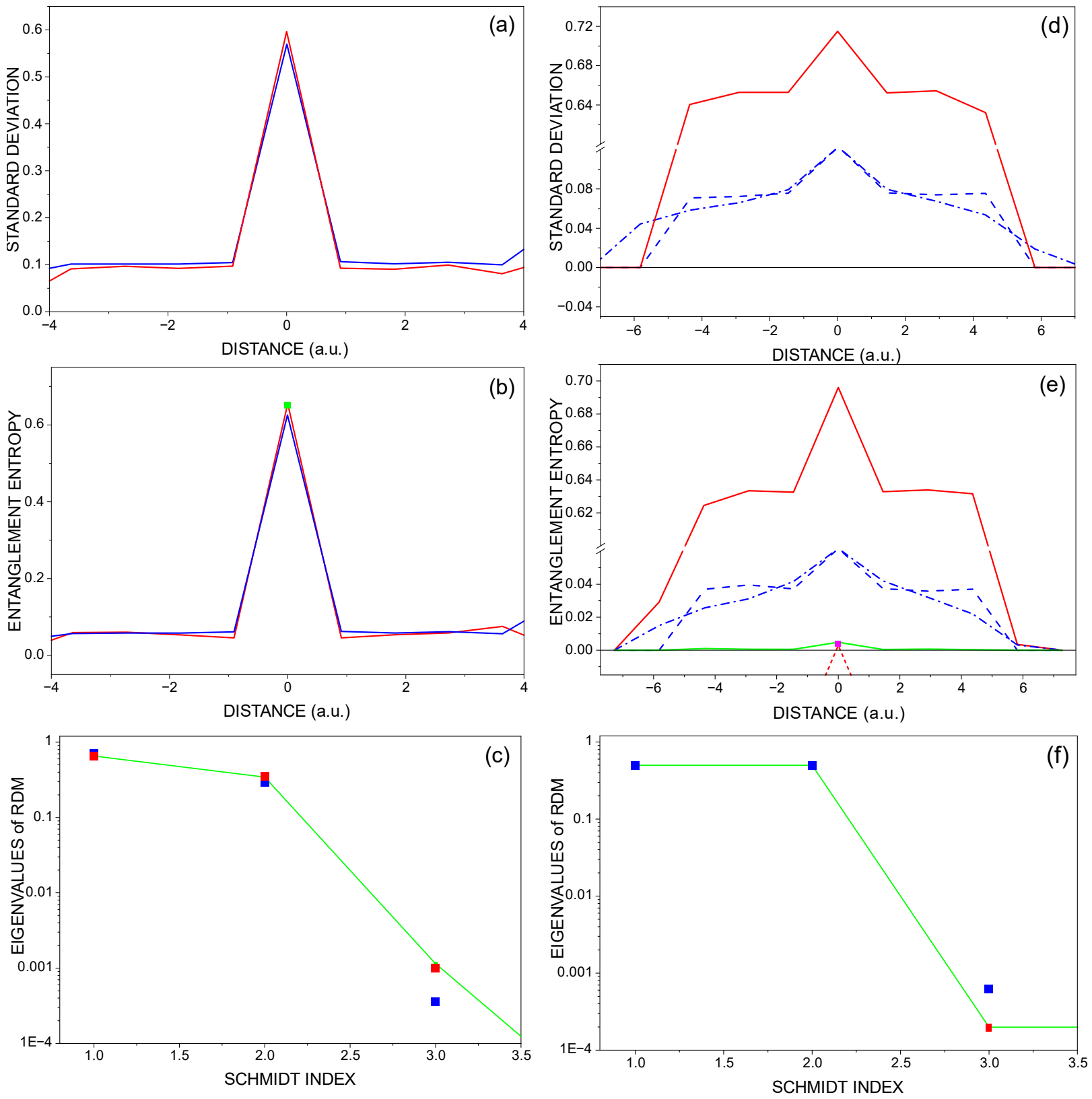

**Figure 5.** Same as Figure 3, but for the one-dimensional hydrogen molecule. Panels (a–c) correspond to parahydrogen and panels (d–f) to orthohydrogen. All color codes, line styles, markers, Schmidt-spectrum conventions, and ln(2)-corrected entropy curves are defined as in Figure 3.

For orthohydrogen, the spatial profiles of the functional standard deviation and entropy (Figure 5(d,e)) resemble those observed in the atomic case, reflecting the similarity of the underlying walker distributions for fermions. As before, subtracting ln(2) from the exact local fermionic entropy leads to negative values (red dashed line in Figure 5(e)), whereas the TDQMC results remain positive and operationally well-defined across all domains (green line). The corresponding Schmidt spectra (Figure 5(f)) confirm that the TDQMC ensemble accurately captures the effective dimensionality of the system.

It is also worth noting that the quantitative agreement between TDQMC and the exact results is less pronounced for orthohelium and orthohydrogen than for the corresponding opposite-spin systems. A possible explanation is that the local entanglement structure of parallel-spin states is strongly influenced by exchange-induced nodal constraints. Since TDQMC incorporates exchange statistically through the ensemble of guide waves rather than through an explicitly antisymmetric many-body wavefunction, some quantitative deviations from the exact local entropy profiles are expected. Nevertheless, the principal maxima and minima occur in similar spatial regions, indicating that the dominant localization patterns of entanglement are still captured.

Finally, we note that the computation of the Gram matrix involves pairwise overlaps between marginal wavefunctions. Enhanced entropy arises when multiple marginal wavefunctions contribute coherently within a spatial region. This reflects the buildup of genuine quantum correlations associated with electronic delocalization, independent of any specific mode decomposition. Notice that for large numbers of walkers, it may still be computationally more efficient to evaluate entanglement and coherence measures using the reduced density matrix rather than the Gram matrix.

## 4. Conclusions

The framework developed in this work shows that, within the TDQMC approach, entanglement can be interpreted in terms of the statistical structure of an ensemble of marginal wavefunctions in Hilbert space. In this representation, the Gram matrix of wavefunction overlaps serves as a covariance operator that encodes the second-order statistical properties of the ensemble. The Gram matrix and the reduced density matrix constructed from the same ensemble share identical nonzero eigenvalues; applying this correspondence to TDQMC marginal-wave ensembles provides a direct connection between ensemble statistics and entanglement measures. This connection enables the characterization of entanglement without explicit construction of the many-body wavefunction.

A central result of this work is that the functional standard deviation of the marginal-wave ensemble, which quantifies its Hilbert-space dispersion, exhibits a spatial dependence that closely follows that of the von Neumann entanglement entropy when both quantities are evaluated locally through spatial partitioning of the walker ensemble. Across the model systems considered, both measures display aligned maxima and comparable spatial localization, indicating that regions of enhanced entanglement correspond to increased diversity of marginal wavefunctions. For fermionic systems, the TDQMC framework provides a positive, well-defined way for evaluating spatially resolved entanglement. In approaches based on strict conditional wavefunctions, subtracting the geometric contribution associated with antisymmetry for fermions may lead to negative local entropies over extended spatial regions. In contrast, the TDQMC guide-wave ensemble yields local entropy profiles that remain positive across all domains and exhibit spatial structure consistent with the global entanglement scale. This behavior reflects the fact that TDQMC evolves in physical space and encodes exchange effects statistically at the level of the ensemble.

More generally, the present results support the interpretation of entanglement as arising from the effective dimensionality of the marginal-wave ensemble in Hilbert space. Regions in which more diverse marginal wavefunctions contribute significantly correspond to locally increased mixedness of the reduced density matrix, while regions dominated by a single marginal wavefunction approach a rank-one structure and exhibit weaker entanglement. In this sense, entanglement is associated with the diversity of marginal waves rather than with spatial overlap alone.

The proposed framework provides a direct and computationally efficient route for analyzing spatially resolved quantum correlations without relying on explicit Schmidt decompositions or reconstruction of many-body states. Because it operates in terms of ensembles of one-body wavefunctions coupled to particle trajectories, it is naturally suited to systems in which spatial structure and dynamical correlations play a central role, including low-dimensional systems, nanoscale structures, and time-dependent processes.

Because the Gram matrix is mathematically equivalent to a kernel matrix in the reproducing kernel Hilbert space spanned by the marginal waves, the rich choice of kernel methods from machine learning, such as Nyström approximations, becomes directly applicable. For example, in large TDQMC simulations, the Gram matrix scales as $M \times M$, where $M$ is the number of walkers. Kernel-PCA methods may therefore be used to identify a low-dimensional subspace spanned by the dominant Schmidt modes of the marginal-wave ensemble. Likewise, Nyström sampling could approximate the leading spectrum of the Gram matrix using only a subset of walkers, reducing memory and computational costs while preserving the dominant entanglement structure. Several extensions follow naturally from the present formulation, including spin-resolved Gram matrices for analyzing spin-dependent correlations and applications to explicitly time-dependent regimes for tracking correlation dynamics. Despite the relatively large walker and guide-wave ensembles used in the present calculations, the intrinsically parallel structure of TDQMC enables efficient computation. The ground states of the model atom and molecule considered here can be obtained within a few minutes on a standard multi-core workstation.

**Acknowledgment:** This research is based upon work supported by the Bulgarian Ministry of Education and Science as a part of National Roadmap for Research Infrastructure, grant D01-102/26.06.2025 (ELI ERIC BG), and by the National Science Fund, grant КП-06-Н78/6.